# Super Event Driven System OOP GUI Design

Ehab Aziz Khalil (Ph.D., Senior Lecturer) &  Mahmoud Samir Fayed (UG Student)

December 15, 2005

*Abstract*—This article presents a new proposal design of GUI and new technology in programming Namely 'Super Technology" which can be applied for supporting the proposal design of GUI

*Keywords*— Object Oriented Programming (OOP), Graphical User Interface (GUI), Event Driven System, Embedded System, and new proposal design of GUI.

## I. INTRODUCTION

T he Graphical User Interface, or "GUI", is a computer  interface that uses graphic icons and controls in addition to text. The user of the computer utilizes a pointing device, like a mouse, to manipulate these icons and controls. This is considerably different from the command line interface in which the user types a series of text commands to the computer. The most common GUI systems in world today is Microsoft Windows, Mac OS X & Xwindow system (for linux), see Fig. 1 and Fig. 2.

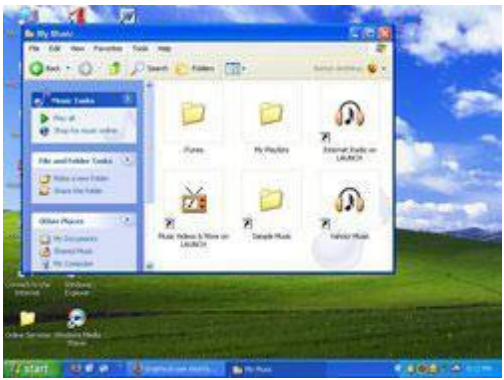

Fig.1 An example of graphical user interface in Microsoft Windows XP

The proposed Super event driven system object oriented programming graphical user interface design is very easy user interface design that achieve all of your dreams about designing stable and powerful GUI programs for any

E. A. Khalil (Supervisor) is with the Dep.t of Computer Science & Engineering, Faculty of Electronic Engineering, Minufiya University, Menouf – 32952, EGYPT (corresponding author phone: 2048-2223906; fax: 2048-2226454 ; e-mail: Ehab_Khalil@mailer.eun.eg, drehab_khalil@yahoo.com or dr.ehab@mailer.menofia.edu.eg ).

M. S. Ibrahim (Author) is with the Faculty of Electronics Engineering Faculty (Minufiya University, Menouf – 32952, Egypt) , msfclipper@hotmail.com .

operating system's environment (such as DOS, Windows, Linux,…, etc.) .

The user of this design will need little effort and time without: involve of either the system's Operating System (OS) event driven system or the interface classes from the programming language because here in our proposal, we are building the GUI environment from the bottom to the top (or from A to Z ).

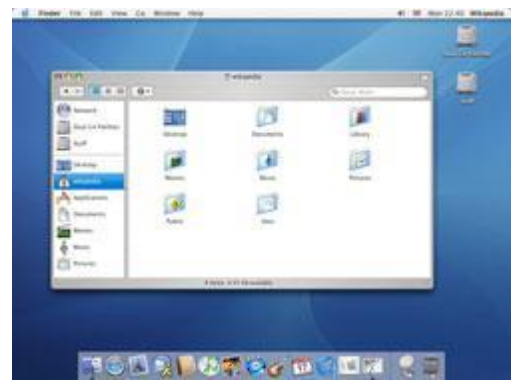

Fig.2 An example of graphical user interface in Apple's Mac OS X

Our proposal gives three advantages; the first one is to develop of the earlier OS's applications like DOS applications, in which we can appearance (view ) of the Window Screen Application (event driven system with GUI) with the same features but more stable and more easy to use.

The second advantage is the ability to the users to have several new general features and multi purpose technologies in programming in different systems (not necessary GUI) that's to improve its work (Stability, Fast and Quality), those technologies provide us with the opportunity to treat a very difficult system in short time with high stability level and clear code easy to learn and modify.

The third advantage is that when the user prepares something with his hand (using easy way) guaranteed to get the best performance which he wants and there is no need to wait for fixing the bugs from other programmers or waiting for adding new features to the product while the user can add it in very short time with little effort.

It's a fact that for having complete GUI environment the system must have embedded **event driven system** which makes simulation for **multi tasking** where the system can be able to receive events while it looks after other jobs in the background.

A certain amount of insight into GUIs can be obtained by comparing noun- verb to verb- noun metaphors. Noun-verb interaction begins by picking an object then telling the system what to do to it. Verb-noun systems tell the system what to do, then pick the object to do it to. Most GUIs are implemented in terms of an event model, although other models exist. These alternative models for creating GUIs are generally classed as User Interface Management Systems or UIMS.

## II. DEFFINITION OF SUPPER TERM

It's well known that the current system in programming supports three techniques:
- new event driven system (Events current circuit)
    - It's new OOP event driven system which use electrical circuits theory for controlling the order of calling events and gives a lot of other features which make multi tasking programming more easy and powerful so it can be used in developing the most complex applications in the world today.
- new client-server system (veto system)
    - This system let user application works as server for other applications where it can receive instructions from other applications and this veto system comes with very powerful rules for controlling this job.
- new multi usage data structure (chemical system)
    - The chemical system is the best data structure for developing the most complex applications and it's come with a lot of features which let it work more than data structure, for example user can use it as (messages management system with OOP, Event driven system, arguments management system & virtual DBMS).

*1) Note:*
It's necessary to mention here that the term (visual) is given to systems that contain:

GUI system. Event
driven system. OOP
style.

Similarly in Visual basic, Visual FoxPro, Delphi, ….etc.

In the proposal design the term (**Super**) to systems that contain :

Events current circuit system.
Veto system.
Chemical system.

## III. SUPER GUI TOOLS

Any user looking for building his own GUI system, he will require some tools for doing that. So, it is necessary to him to know what are the requirements needed before starting building his own GUI system. The requirements are high level programming language which included:
   1 - OOP (Object Oriented Programming) style.
   2 - Graphic functions for drawing only.
   3 - Relational DBMS (DataBase Management System).
   4 - Protected mode linker.
   5 - Good IDE (Integrated Development Environment).

The testing of algorithms with the proposal design by using CA-Clipper, Class(y), Fglib, Xmate and Blinker, yield promising enough results.

**CA-clipper 5.2e** (Computer Associates International company): is a high level language with very powerful database management system (DBMS) from computer associates international (1986 to 1996) and it's work in DOS mode but can be used with some libraries like Clip-4-win for developing windows applications , CA-Clipper is developed using Microsoft C , the last version is CA-Clipper 5.3b but the most stable version is CA-Clipper 5.2e.

**Class(y) 2.4**: its object oriented extension library for CA-Clipper (freeware).

**Fglib 3.1** (Ferns graphics library): it's very nice freeware graphic library for CA-Clipper by Ferns Paanakker, the last version is 3.1 (appear in 2002) but there are new version will coming soon by (Ferns Pannakker & us ) and this new version will contain complete GUI package (as a result of this research) and the version will be Fglib4.

**Blinker 7**: it's shareware linker which can produce DOS exe files (real-dual and extended mode) and can also produce windows exe files, and this linker support static and dynamic DLL (Dynamic Link Library) files , and this linker can be used with more than one programming language like clipper, c++, force ,….etc.

**Microsoft Xmate 1.9z6**: it's freeware project manager which support more than one programming language like CA-Clipper, C++, Etc. (2003-2005) by Andrew Wos.
.

## IV. GUI APPLICATIONS

1 - We can use GUI system for developing normal applications (DOS mode).
2 – We can use GUI system as embedded system in OS (Operating System).
3 – We can use GUI system as embedded system in PL (Programming Language).

**Note that** : It's well known that the age of DOS is almost expired, but there are now free DOS which is freeware DOS version (32bit & multi tasking) and can be used as platform for DOS applications that does not need GUI environment like windows for example IC (Integrated Circuits) interface applications which used in factories with PIC (Programmable Integrated Circuits). IC interface applications need multi tasking and event driven system user interface (text mode or graphic mode) , the user can now let it work in GUI environment away from shareware OS like Windows using this research.

### A- GUI design features:

For having great GUI management system design it should included the following:
- Contains simple interface tools,
- Nice help guide for user,
- Powerful event driven system,
- Interface objects should support mouse and keyboard in the same time,
- Powerful screen unit support layers system for windows and redraw system ,
- Good skin through color programming theories,
- Using OOP style for writing code,
- Supporting windows system where we can move between windows,
- Full control on the appearance of the user interface,
- Support dynamic link library files so multiple programs can share the GUI,
- Intelligence in memory management,
- The system should work very fast,
- High level of Stability is required,
- The time of the design application process should be little,
- The system should be open architecture (extended),
- The design should be not complex.

### B- Other user programmer:

The windows programmer may ask "What are the helps of the new proposal ?"
The following points are answering that question :
- This design for making the user own event driven system (easy, stable & powerful) using Super technology (ECS,CS,VS),
- The games programmers can make the user's special GUI management system using this research plus directX,
- The user can build this own GUI using this design so that he can add more classes that present new objects with specific properties that he wants,
- The user can create new system which work under windows and has own GUI management system for more control and stability,
- The user can create new programming language under windows that contain embedded GUI management system.

### C-. Motivation:

It's a fact that the programming world grow very fast and every day we are seeing new technologies some of these may be utilities programs, classes, libraries or new programming languages, so, obviously it's very dynamic world and it's coming with very strange features where some times through the application you can't determine if the programmer is professional or not.

For example you may see very nice application with powerful GUI and if you ask the programmer to change the appearance of the interface may be he can't because, he use static libraries, tools and classes.

It's well known that the tools consume the time but it should also not consume the power, we should be aware of the tools in our world to achieve the most progress in little time but in specific domains we should also know the theories and the algorithms of the material to have the ability to build, improve and develop any required result.

So the ability to make an urgent GUI with specific features and properties that reserve our system is very necessary for every programmer to be adaptable with market changes for that we start to build the Super GUI research to present the algorithm and design of the GUI for every programmer with simple way and powerful design easy to learn, use and develop.

In this article there is a proposal for new technologies in programming to preserve the time.

### A- Related Works:
#### 1- Examples:
- Microsoft windows interface is very powerful and the stability is not bad but every application work under windows use it's event driven system and it's graphical users interface and this gives us some stability problems sometimes and for best performance we should build our own event driven system and graphical user interface classes using this technology and develop it as you can to match your needs.
- Most DOS applications before 1995 don't support movable windows system.
- Dual DOS applications that work in text mode and graphic mode gives us bad GUI like CA-CLIPPER 5.3 applications.
- Very strange and common drawback in GUI is giving user object's that work with mouse only (button for example) and there is no support for mouse or giving the user objects that work with keyboard only (textbox for example) where there is no support for mouse.
- Some GUI applications don't support mouse event's very good (mouse move events).
- Some GUI don't have event driven system (very weak environment) or not OOP so it's difficult environment.
- Some GUI don't support VESA driver to access Hicolor modes and it is graphic library problem (you will not face this problem in windows).

*2- SEAL 2.00.11 program:*

Copyright © SEAL Developers 1999-2002.
All Rights Reserved.
Web site: http://sealsystem.sourceforge.net/
E-mail: orudge@users.sourceforge.net

"SEAL is a 32-bit graphical user interface for DOS. It supports high-resolution display modes, multimedia, and more! . The source code is freely available and you are encouraged to help with SEAL in whatever way you can - whether you can program or not, every little helps. "
README.TXT from seal files

- The seal program is nice but it's GUI is not very powerful yet because
    i. There is no support for keyboard in command buttons and icons
    ii. The mouse event's is not very good (mouse up and mouse move events)
    iii. It's not OOP
    iv. I test it in win98 and it's work good in Hicolor but when I test it in winXP it's not work and it's graphic library problem

- At the end I would like to say that the seal program has very nice movable windows management system and I use it sometimes when I return to the old world of DOS.

*3- NeoPaint(R) for DOS - Version 3.2 (C) 1992-97 NeoSoft Corp:*

ABOUT NEOPAINT: "Simply one of the best paint programs you'll find. NeoPaint continues to receive rave reviews from numerous publications, and has quickly become one of the most popular DOS paint programs in the world. NeoPaint is extremely easy to use, yet includes features which would cost hundreds of dollars more with other programs. NeoPaint's abundance of drawing tools makes it easy to create illustrations for desktop publishing projects, edit digitized photos or just explore your artistic side. NeoPaint reads and writes 2, 16, 256 and 16 Million color PCX, TIFF, BMP and GIF files, includes dozens of drawing tools, special effects, stamp pad and more."
README.TXT file from NeoPaint files

- It's very nice GUI application for DOS.
- We don't have the source code to determine if it's structure programming or OOP.
- The movable windows system is good but need some support for moving form window to other during runtime.
- There are standard font used for GUI objects and I really don't know if it's easy to change the font scale & type or this need big effort from the programmer if the classes is designed to work on one scale.
- The application appearance is standard and there is no hicolor skin.
- The application is not user friendly very well for example there is no tool tip text for the graphical toolbar.

4- *Microsoft Windows XP:*
   a. It's very powerful operating system
   b. Comes with great skin.
   c. It's suffer from some stability problems when you run a lot of big applications in the same time due to time response problem.
   d. Using the proposal design we can improve the Microsoft windows XP event driven system where we can have more control on the order of events at run time.
   e. Using this proposal design the we can make Microsoft Windows XP event driven system more easy to learn and use.
   f. Using this proposal design we can make Microsoft windows XP GUI is great client-server windows management system using veto system.

*5- Real design (45 classes):*

The proposal GUI design contains (event driven system, windows management system and GUI widgets). And it's designed by using OOP style and the system contains 45 classes, see Fig. 3 for the proposal design block diagram. From Fig.3 we come to know that the GUI system contains two components the first one is graphic library for drawing only (kernel) and the second one for event driven system, windows management system and GUI widgets (surface) and tour proposal design concerned with the second component.

The windows management system is responsible for supporting work with multiple windows (movable and sizeable) and also responsible for redrawing screen so the proposal design comes with screen unit with two systems (layers system and redraw system.). The proposal design supports mouse and keyboard simultaneously, so there are mouse unit, keyboard unit and focus system for moving focus automatically between objects using mouse or special keys like arrows, Tap and Shift+Tap.

One of the most application of the GUI is the Desktop, so the proposal design presents desktop classes for supporting icons, taskbar and nested level menus. The proposal design presents the main GUI widgets required for any GUI application like (Label, Shape, Image, Button, textbox, editbox, listbox, scrollbar, compobox, frame, checkbox, pages, progress bar, menu bar,….etc). The event driven system design (events current circuit) which used in the GUI management system design in this proposal design is a part of the Super technology. The proposal design contains the veto system which let the GUI system work as server for other applications at run time or for controlling the response level of messages between objects in the same application, see Fig 4 which show the block diagram of the veto system, and see Fig 5 which show the theory of veto system. The proposal design comes with chemical system (one part of Super technology) for adding more features to the design where the chemical system is new data structure but it's more than that because it can be used as event driven system, arguments management system and virtual relational database, see Fig.6 which show the chemical system block diagram and note that the (electron) may be a (variable, array, object or any random data).

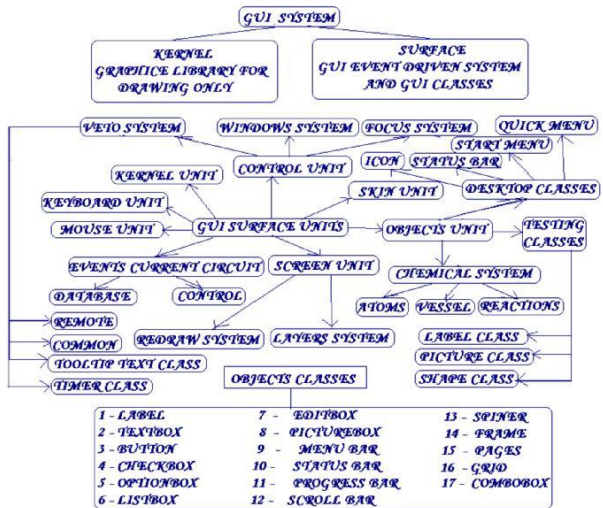

Fig.3 Proposal Design Block Diagram.

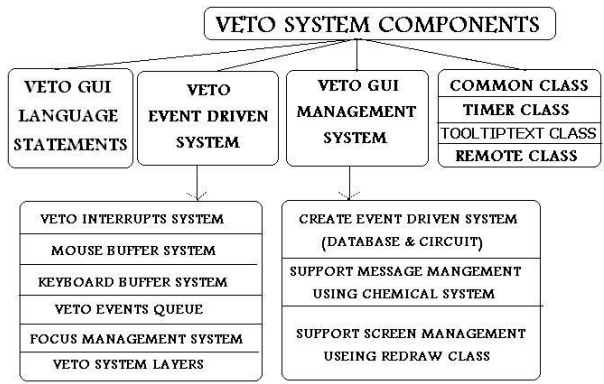

Fig.4 Veto System Block Diagram.

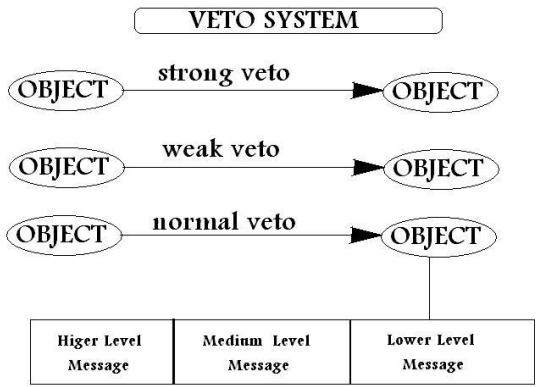

Fig.5 Veto System Theory

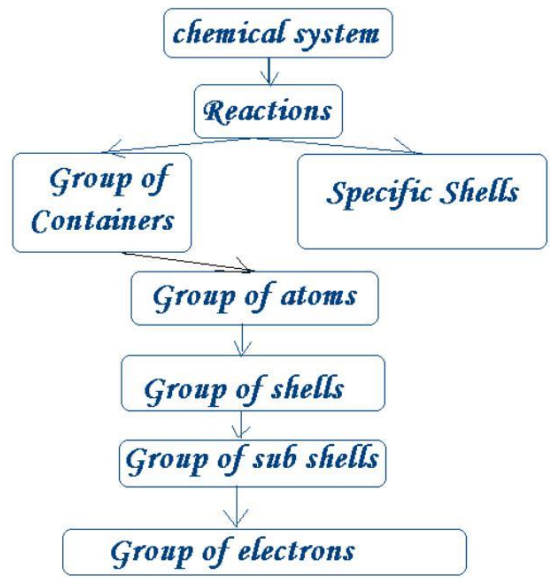

Fig.6 Chemical System Block Diagram

## V. COMMON CLASS THEORY

Fig.7 shows the common class usage.

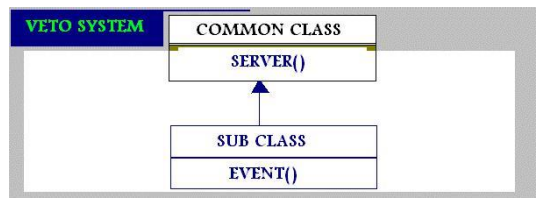

Fig.7 Common Class Usage

In the world of OOP the inheritance gives us very nice advantage to preserve us from the repetitions of code so we should use it of course.
In the GUI world the application contain more than one object from different classes but these classes have common features. In event driven system we have group of constant mouse and keyboard events which we deal with them.
The common class is a class that handles the common events for us that we need In each GUI class.
If we want to make GUI class like BUTTON CLASS we can inherited it from the common class to reduce more than 50% of our development time.
The common class have the SERVER() method which determine the mouse or keyboard event from (20 different event).

*Server Method Events:*

*Table:1 shows the default events handled automatically by common class.*

| NUM. | KEYBOARD OR MOUSE EVENT |
|---|---|
| 1 | KEY PRESS EVENT |
| 2 | SPECIFIC KEY PRESSED |
| 3 | NOT REQUIRED KEY PRESSED |
| 4 | MOUSE IN REGION |
| 5 | LEFT MOUSE BUTTON DOWN |
| 6 | RIGHT MOUSE BUTTON DOWN |
| 7 | MOUSE BUTTON DOWN |
| 8 | LEFT MOUSE BUTTON CLICKED |
| 9 | LEFT MOUSE BUTTON DBCLICKED |
| 10 | LEFT MOUSE BUTTON UP |
| 11 | RIGHT MOUSE BUTTON CLICKED |
| 12 | RIGHT MOUSE BUTTON DBCLICKED |
| 13 | RIGHT MOUSE BUTTON UP |
| 14 | MOUSE MOVE EVENT |
| 15 | MOUSE MOVE AND LEFT MOUSE BUTTON CLICKED OUT |
| 16 | MOUSE MOVE AND RIGHT MOUSE BUTTON CLICKED OUT |
| 17 | MOUSE MOVE AND MOUSE BUTTON CLICKED OUT |
| 18 | LEFT MOUSE BUTTON DOWN OUT |
| 19 | RIGHT MOUSE BUTTON DOWN OUT |
| 20 | MOUSE BUTTON DOWN OUT |

Table:1 Default Events Handled Automatically by Common Class.

## VI. SIMULATION RESULTS

The real results of this application are complete GUI management system which contain the following features: (see Fig.8- Fig.14)

Super programming technology features (chemical system, events current circuit & veto system).

Complete multi tasking GUI environment where the GUI can respect events from user through mouse and keyboard while it do other jobs in back ground.

Very powerful windows management system which support dealing with multiple windows in the same time and you can (move, size, maximize & minimize) any window.

The system contain desktop classes (icon, nested level menus & taskbar)

The system contain very powerful focus system which support mouse and keyboard in the same time

The system contain layers system & redraw system for screen management

The system contain nice skin (256 color & hicolor)

The system contain all standard GUI widgets (button, textbox ,…..etc)

The system contain classes for supporting creating new GUI widgets

The system contain nice Form Designer utility

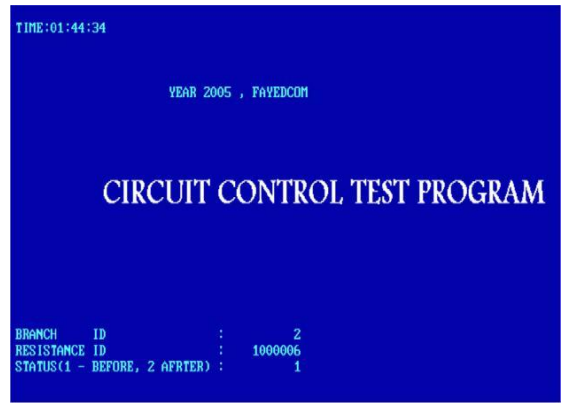

Fig.8 Event Driven System (Events Current Circuit) Test Program.

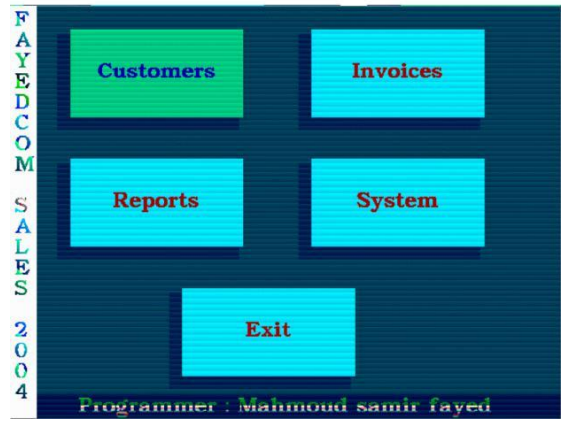

Fig.9 Skin Unit Test Program.

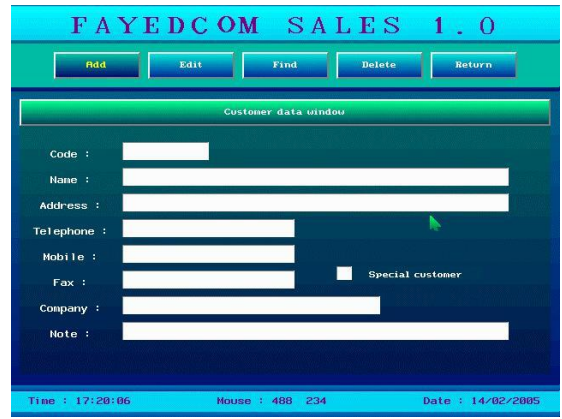

Fig.10 Focus System Test Program.

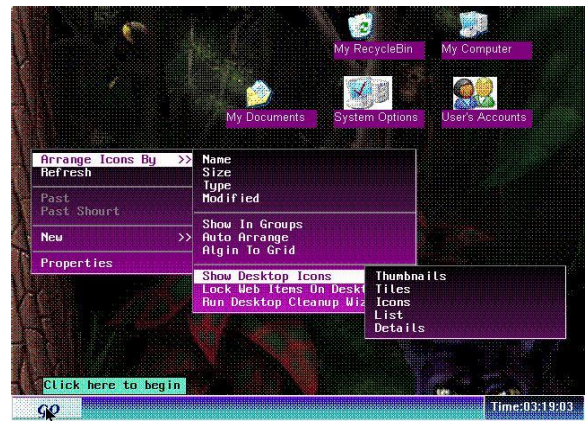

Fig.11: Nested Level Menu Test Program.

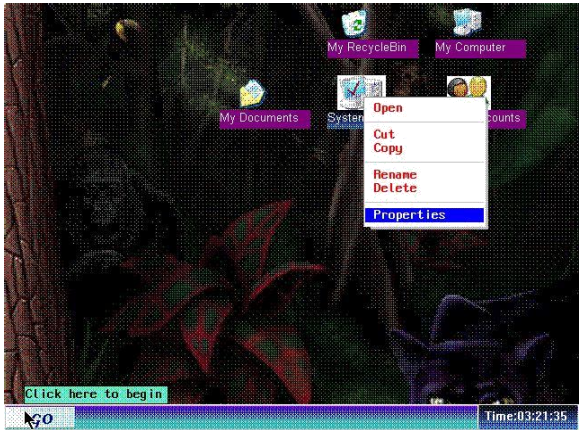
Fig.12 Desktop Test Program.

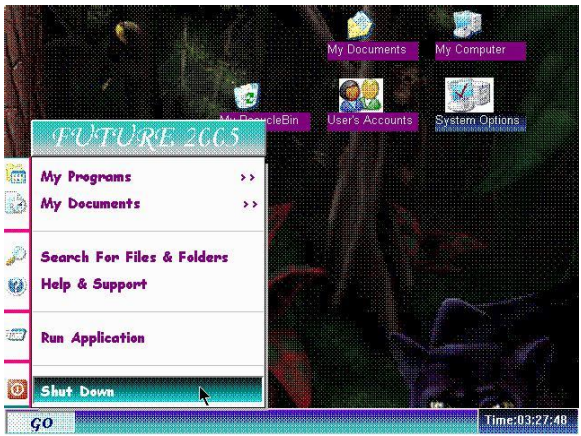
Fig.13: Task Bar Test Program.

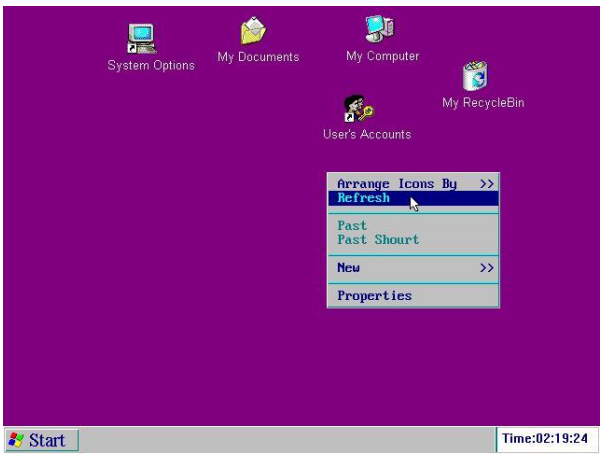
Fig.14 Desktop Test Program (without skin).

## VII. CONCLUSION

The proposal design of GUI which is promising enough with the current running programming, that is clear from the wide band of applications with a few and easy code. The proposal design of GUI includes very high level of technology such as powerful event driven system, chemical system, veto system, OOP, Relational DBMS and Hicolor skin.

The proposal design will be very helpful and useful for all levels of programmers which are interested for improving and increasing their products with the new information technologies.

Future work:

The proposal design of GUI as well as the super programming technology design guided us for creating a new programming style Vs OOP.

The new programming style will completely take place of OOP, if that success that means the programming world will be definitely change.

**Dr. Ehab A. Khalil, (Senior Lecturer)** (B.Sc'78 – M.Sc.'83 – Ph.D.'94), Bourn in Sohage – Government - EGYPT, on July 5$^{th}$ , 1955, Got B.Sc. in the Dept. of Industrial Electronics, Faculty of the Electronic Engineering, Minufiya University, Menouf – 32952, EGYPT, in May 1978, M.Sc in the Systems and Automatic Control, with the same Faculty in Oct. 1983, Research Scholar from 1988-1994 with the Dept. of Computer Science & Engineering, Indian Institute of Technology (IIT) Bombay-400076, India, Ph.D. in Computer Network and Multimedia from the Dept. of Computer Science & Engineering, Indian Institute of Technology (IIT) Bombay-400076, India in July 1994. Lecturer, with the Dept. of Computer Science & Engineering, Faculty of Electronic Engineering, Minufiya University, Menouf – 32952, EGYPT, Since July 1994 up to now. Participated with the TCP of the IASTED Conference, Jordan in March 1998. With the TPC of IEEE IC3N, USA, from 2000-2002. Consulting Editor with the "Who's Who?" in 2003-2004. Member with the IEC since 1999. Member with the Internet2 group. Manager of the Information and Link Network of Minufiya University, Manager of the Information and Communication Technology Project (ICTP) which is currently implementing in Arab Republic of EGYPT, Ministry of Higher Education and the World Bank. Published more than 70 research papers and article reviews in the international conferences, Journals and local newsletter.
For more details you can visit http://ehab.a.khalil.50megs.com or http://www.menofia.edu.eg/network_administrtor.asp and to contact please use: Ehab_Khalil@mailer.eun.eg, dr.ehab@mailer.menofia.edu.eg, drehab.khalil@gmail.com, drehab_khalil@yahoo.com, khalilehab@hotmail.com.

**Mahmoud Samir Fayed (UG Student)** Faculty of Electronics Engineering Faculty (Minufiya University, Menouf – 32952, Egypt) Well use of C++, CA-Clipper & Visual FoxPro Programmer, interested in Database Programming & Embedded Multi Tasking GUI systems designing For contacting, msfclipper@hotmail.com.